\documentclass{kluwer}    % Specifies the document style.

\usepackage[]{graphicx}
\newdisplay{guess}{Conjecture}

\begin{document}                                                                                   
\begin{article}
\begin{opening}         
\title{Monitoring and modeling radio flares from microquasars}
\author{Sergei \surname{Trushkin}}      %\email{satr@sao.ru}
\author{Elena  \surname{Majorova}}
\author{Nikolai \surname{Bursov}}
\runningauthor{Trushkin et al.}
\runningtitle{Radio flares from microquasars}
\institute{Special Astrophysical Observatory of Russian AS}
\date{October 15, 2000}

\begin{abstract}
We present results of long-term daily monitoring of a sample
of Galactic radio-emitting X-ray binaries showing relativistic jets
(RJXRB): SS433, Cyg~X-3, and GRS~1915+105, with
the RATAN-600 radio telescope in the 0.6--22~GHz range.

We carried out the modeling calculations to understand the temporal
(1--100 days) and spectral (1--22~GHz) dependence.
We tested the finite jet segment models and we found that the geometry of
the conical hollow jets is responsible for either a power law or an
exponential decay of the flares.

SS433 was monitored for 100 days in 1997 and 120 days in 1999.
From the quiescent radio light curves, we obtained clear evidence
of a 6.04-day 10--15\% modulation. 

Three powerful flares (up to 13~Jy) from Cyg X-3 were
detected in April 2000.
\end{abstract}

\end{opening}           

\section*{Results of the multi-frequency daily monitoring}
Many (or all!) of the radio sources related with X-ray binaries
have been resolved into relativistic jets, often with detectable
proper motion of the radio-emitting blobs.
We present results of long-term daily monitoring of a sample of these
RJXRBs with the RATAN-600 radio telescope at 0.96, 2.3, 3.9, 7.7, 11.2 
and 21.7~GHz.
Many strong optically thin and thick flares were detected during the 
monitoring period (see Bursov and Trushkin, 1996; Trushkin, 1998).
The basic model of the flares is the synchrotron emission evolution of two
relativistic jets with conical geometry and an adiabatic expansion of
the blobs, which are composed of relativistic electrons embedded in 
magnetic fields, moving away from the central source (Shklovskii,
1960; van der Laan, 1966; Hjellming and Johnston, 1988; Marti et al.,
1992). The flare decay law is determined by the geometry of the conical
hollow jets. The jet velocity, thermal and relativistic electron densities,
and magnetic field intensity within the blobs can be derived from the fits.

\begin{figure}
\centerline{\hbox{\includegraphics[width=6.2cm]{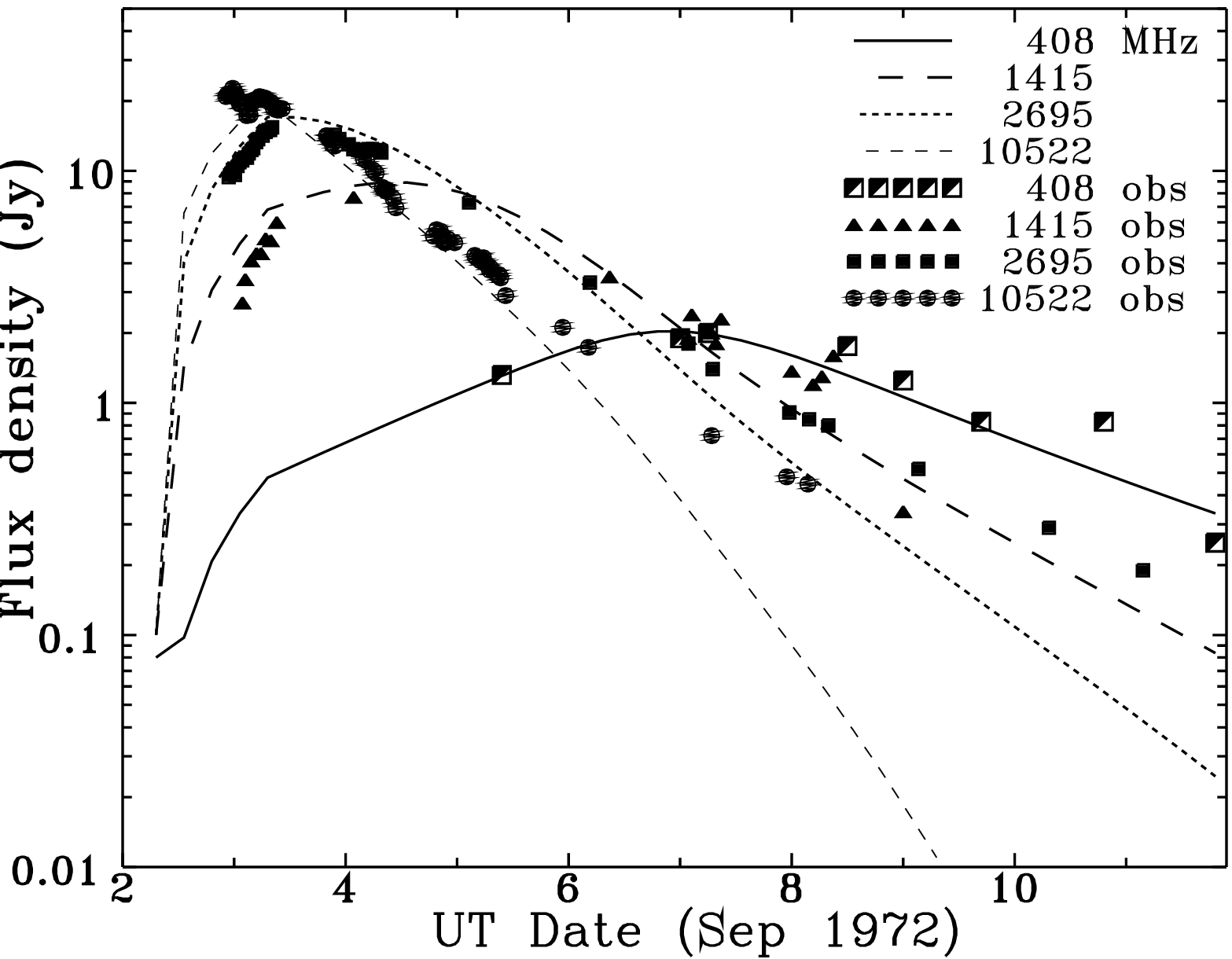}
\includegraphics[width=5.5cm]{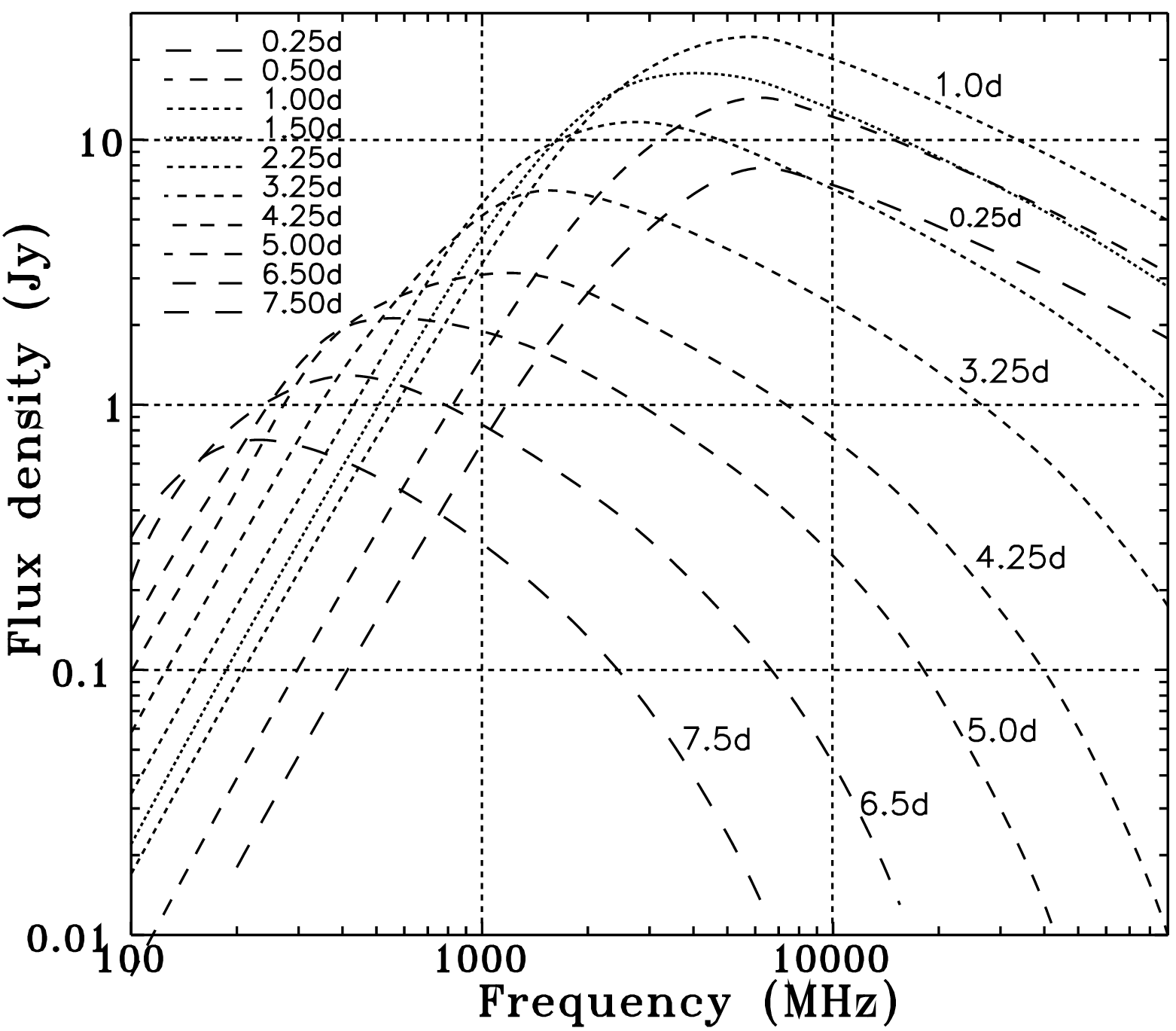} }}
\caption{The model light curves at select frequencies
(left) with the real data and the spectra
of the famous Cyg~X-3 flare at 2--11 September 1972 (right).
The model parameters are from Marti et al. (1992)}
\label{Marti}
\end{figure}

Our monitoring of radio variability of RJXRB shows that the decay of a
flare follows a power law, $S_{\nu} = S_0 t^{-2p}$, as predicted
by the Shklovskii and van der Laan models, or an exponential law
$S_{\nu} = S_0 e^{-t/\tau}$ (Fig.1), which is due to the
initial exponential increase of the jet cone radius.
Often $\tau \sim \nu^{\beta}$, where $\beta$
ranged from $-0.8$ to $-0.4$; these flares decay faster at higher 
frequencies (see current light curves and the spectra animation at \verb*Chttp://cats.sao.ru/~satr/XB/C).

We computed the power spectra of the daily light curves of SS433 in the
2.3--22~GHz range during the long quiescent periods May--July 1997 and
May--August 1999. Both spectra show
a mean harmonic at $0.165\pm0.02 \ {\rm day}^{-1}$,
which corresponds to a period of $6.04\pm0.06$ days.

We propose that Doppler boosting of the jet radiation could explain
such modulation. The 5-elements kinematical model of the jets and
ephemerides proposed by Vermeulen (1989) for the nodding motion of
the jets and  common formulas for boosting produce a modulation similar
to the observed one. Thus we detect a 6-day modulation at the 10--15\%
level in the quiescent flux density, which is
$S_{\nu}[{\rm Jy}] = 1.2 \ \nu_{\rm GHz} ^{-0.6}$.

Three powerful flaring events (up to 13~Jy) from Cyg X-3 were
detected in April 2000.
The daily spectra clearly show two components: i) a
quasi-stable steep spectrum and ii) a very unstable flat spectrum.

In the wide range 0.6--22 GHz the optically thin spectrum
of the detected strong flare of GRS~1915+10 (5 August, MJD~51751.85)
is well fitted by the power law:
S$_\nu$[Jy] = 0.690\,$\nu^{-0.43}_{\rm GHz}$.

%\acknowledgements  %  I could not used  -- there is no place!!!
\vspace{0.3cm}
{\small The authors are grateful to  RFBR for supporting the project
of monitoring X-ray binaries, grant N98-02-17577.}

\end{article}

\begin{thebibliography}{}

\bibitem[\protect\citeauthoryear{Bursov and Trushkin}{1995}]{BT}
Bursov, N.N. and S.A. Trushkin
%\newblock{Multi-frequency observations of a recent radio flare in SS433}.
\newblock {\em Pis'ma in Astron. Zh.}, 21, 163--167, 1995.

\bibitem[\protect\citeauthoryear{Hjellming and Johnson}{1988}]{HJ}
Hjellming, R. and K.J. Johnson.
\newblock {\em ApJ}, 328: 600, 1988.

%\bibitem[\protect\citeauthoryear{Katz et al.}{1982}]{K2}
%Katz, J.I., S.F. Anderson, B. Margon, S. Grandi.
%\newblock{\em ApJ}, 260:780--785, 1982.

\bibitem[\protect\citeauthoryear{Marti et al.}{1992}]{Ma}
Marti, J., J.M. Paredes, R. Estalella.
%Modeling Cygnus X-3 radio outbursts: particle injection into twin jets.
\newblock{Astron. Astrophys.}, 258:309--315, 1992.

%\bibitem[\protect\citeauthoryear{Mirabel and Rodr\'\i guez}{1994}]{MR94}
%Mirabel, I.F. and L.F. Rodr\'\i guez.
%% Sources of relativistic jets in the Galaxy,
%\newblock{\em Ann. Rev. A\&A}, 37, in press, 1999.

\bibitem[\protect\citeauthoryear{Shklovskii}{1960}]{Sh}
Shklovskii, I.S.
\newblock{\em Soviet Astronomy}, 4:243--253, 1960.

\bibitem[\protect\citeauthoryear{Trushkin}{1998}]{T3}
Trushkin, S.A.
%Multi-frequency monitoring of the Galactic X-rays sources. I. Cyg X-3.
\newblock{\em Pis'ma in Astron. Zh.}, 24:19--25, 1998.

\bibitem[\protect\citeauthoryear{van der Laan}{1966}]{Laan}
van der Laan, H.
\newblock{\em Nature}, 211:1131--1134, 1966.

\bibitem[\protect\citeauthoryear{Vermeulen}{1989}]{Verm}
Vermeulen, R.C.
%\newblock{\em Multi-Wavelength Studies of SS433}.
\newblock{\em Ph.D. Thesis}. University of Leiden, The Netherlands, 1989.
\end{thebibliography}
\end{document}